\documentclass[aps,prl,twocolumn,showpacs,showkeys,footinbib,superscriptaddress]{revtex4-1}
\usepackage{amsmath}
\usepackage{amssymb}
\usepackage{graphicx}
\usepackage{bm}
\usepackage{color}
\usepackage{relsize}
\usepackage{braket}
\usepackage[caption=false]{subfig}
\usepackage{hyperref}
\usepackage[all]{hypcap}
\usepackage{siunitx}

\renewcommand{\vec}[1]{\mathbf{#1}}

\begin{document}
\title{Optically Probing Tunable Band Topology in Atomic Monolayers}
\date{\today}

\author{Gaofeng Xu}
\affiliation{Department of Physics, University at Buffalo, State University of New York, Buffalo, NY 14260, USA}
\author{Tong Zhou}
\email{tzhou8@buffalo.edu}
\affiliation{Department of Physics, University at Buffalo, State University of New York, Buffalo, NY 14260, USA}
\author{Benedikt Scharf}
\affiliation{Institute for Theoretical Physics and Astrophysics and 
W\"{u}rzburg-Dresden Cluster of Excellence ct.qmat, University of  W\"{u}rzburg, Am Hubland, 97074 W\"{u}rzburg, Germany}
\author{Igor \v{Z}uti\'c}
\affiliation{Department of Physics, University at Buffalo, State University of New York, Buffalo, NY 14260, USA}
\begin{abstract}
In many atomically thin materials their optical absorption is dominated by excitonic transitions. It was recently found that optical selection rules in these materials are influenced by the band topology near the valleys. We propose that gate-controlled band ordering in a single atomic monolayer, through changes in the valley winding number and excitonic transitions, can be probed in helicity-resolved absorption and photoluminescence. This predicted tunable band topology is confirmed by combining an effective Hamiltonian and a Bethe-Salpeter equation for an accurate description of excitons, with first-principles calculations suggesting its realization in Sb-based monolayers.  
\end{abstract}
\maketitle

Since Dirac's magnetic monopoles~\cite{Dirac1931:PRS}, the use of topological quantum numbers has grown significantly and for decades has successfully elucidated fascinating condensed-mater phenomena~\cite{Thouless:1988}. Unlike extensive transport measurements of topological states, starting with quantum Hall effects~\cite{Klitzing1980:PRL,Tsui1982:PRL}, considerably less is known about optically probing tunable band topologies in electronic materials. We suggest that can be changed based on:  (i)  advances in atomically-thin van der Waals (vdW) materials~\cite{Ajayan2016:PT,Reis2017:S} and (ii) realizing that in some of them the formation of excitons, bound electron-hole pairs, also reflects the underlying topological properties~\cite{Zhang2018:PRL,Cao2018:PRL}.

Guided by the spin-orbit coupling (SOC)-driven topological band inversion in HgTe/CdTe quantum wells~\cite{Konig2007:S}, we seek suitable SOC in atomic monolayers (MLs) for a tunable band topology~\cite{Ajayan2016:PT,Reis2017:S}. While graphene has motivated important advances~\cite{Kane2005:PRL,Kane2005:PRL2}, its weak SOC and a negligible gap $\sim$40 $\mu$eV~\cite{Sichau2019:PRL} make its topological properties very fragile. In the opposite limit, ML Bi on a SiC substrate has a strong SOC with a huge topological gap ($\sim0.8$ eV)~\cite{Reis2017:S}, but its indirect gap is unlikely to support excitons and its large characteristic energies make altering the band ordering very challenging.

\begin{figure}[ht]
\centering
\includegraphics*[width=8.6cm]{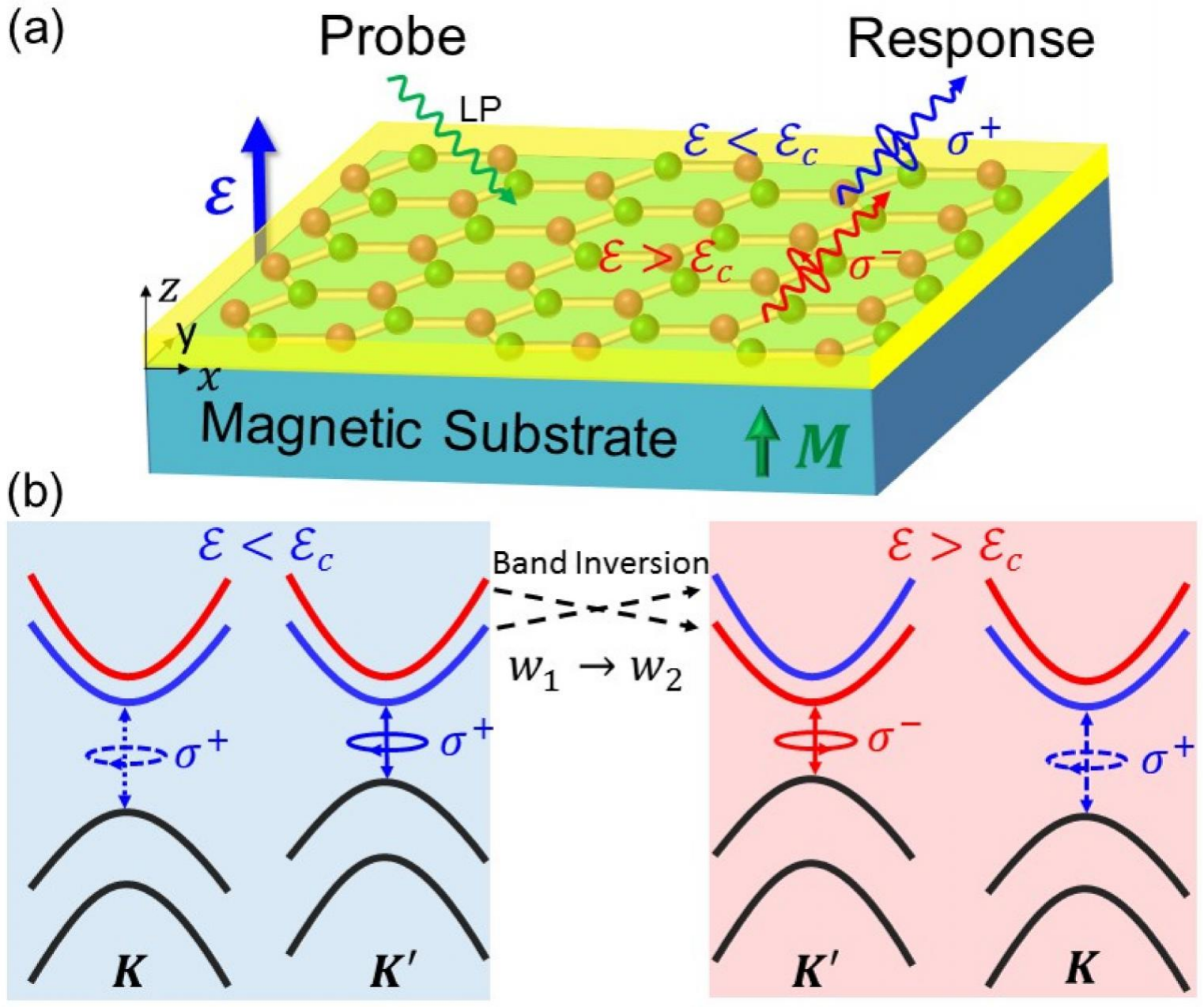}
\vspace{-0.5cm}
\caption{(a) Schematic setup. The linearly polarized (LP) probe light yields a response of definite helicity,  $\sigma^{\pm}$, \,that depends on the electric field, $\mathcal{E}$. (b) Bands at the $K/K'$ valleys. The $K'$-valley conduction band ordering and the winding numbers, $w_{1,2}$, change when $\mathcal{E}$ exceeds a critical value, $\mathcal{E}_c$.
}
\label{fig:Scheme}
\vspace{-0.3cm}
\end{figure}

Instead of these extreme cases, we expect that MLs with intermediate SOC strengths and a direct band gap are more promising. A tunable band topology could be experimentally probed by a change in optical selection rules, as a consequence of the change in the relevant topological numbers. 
Remarkably, we reveal how this topology in a single layer becomes tunable by simply changing an applied electric field in the presence of magnetic proximity effects, shown in Fig.~\ref{fig:Scheme}, or with magnetic doping.

The connection between band topology and optical selection rules was recently established for chiral fermions~\cite{Zhang2018:PRL,Cao2018:PRL}, such as graphene-based systems and transition-metal dichalcogenides (TMDs), where the winding number~\cite{Asboth:2016} characterizing the topology of Bloch bands directly modifies the allowed optical transitions and the excitonic properties. As in the generalization of semiclassical equations of motion to include Berry-phase effects~\cite{Xiao2010:RMP}, recognizing the importance of winding numbers in the optical response opens novel opportunities. 

Unlike previous work on the role of winding numbers on excitons from chiral fermions excluding spin and SOC~\cite{Zhang2018:PRL,Cao2018:PRL}, we consider breaking time-reversal symmetry in proximitized MLs~\cite{Zutic2019:MT} and therefore explicitly include valley and spin degrees of freedom~\cite{Xu2014:NP,Klinovaja2013:PRB}. This provides a direct connection between the helicity-resolved response and the band topology, established both from the single-particle Hamiltonian and with an accurate inclusion of Coulomb interaction in a Bethe-Salpeter equation. Furthermore, we identify how a gate-controlled 
conduction band inversion could be implemented in realistic ML materials 
supporting also a robust topological transport response with nonzero Berry curvature and anomalous valley Hall effect. We reveal that the resulting helicity reversal  
is associated with a sign change of the winding number, rather than its magnitude~\cite{Zhang2018:PRL,Cao2018:PRL}.

Some ML vdW materials, such as TMDs, are known for their optical properties dominated by excitons with large binding energies (up to $\sim0.5$ eV) and efficient light emission~\cite{Wang2018:RMP,Mak2010:PRL}. Gradually, it is becoming understood that their difference from extensive work on excitons in conventional semiconductors is not simply a much larger binding energy, but rather that Berry-phase effects~\cite{Garate2011:PRB,Zhou2015:PRL,Srivastava2015:PRL,Wu2017:PRL,Culcer2017:PRB} and collective excitations can play an important role in the optical response~\cite{VanTuan2017:PRX,VanTuan2019:PRB}.  While we focus on the influence of tunable band topology on excitons, which are absent within the single-particle picture, for an initial characterization of promising materials systems supporting topological states, a single-particle effective Hamiltonian can still offer useful guidance. Group V-based MLs~\cite{Zhang2018:CRC,Gui2019:JMCA}, provide a suitable platform for topological states since the very large SOC and indirect gap of Bi-based MLs~\cite{Reis2017:S} could be altered by considering lighter elements. Specifically, a Sb-based ML, antimonene, 
could support both a smaller SOC and a direct gap~\cite{Note:SM} and has been experimentally realized~\cite{Ji2016:NC,Shao2018:NL}.

\begin{figure}[t]
\centering
\includegraphics*[width=8.6cm]{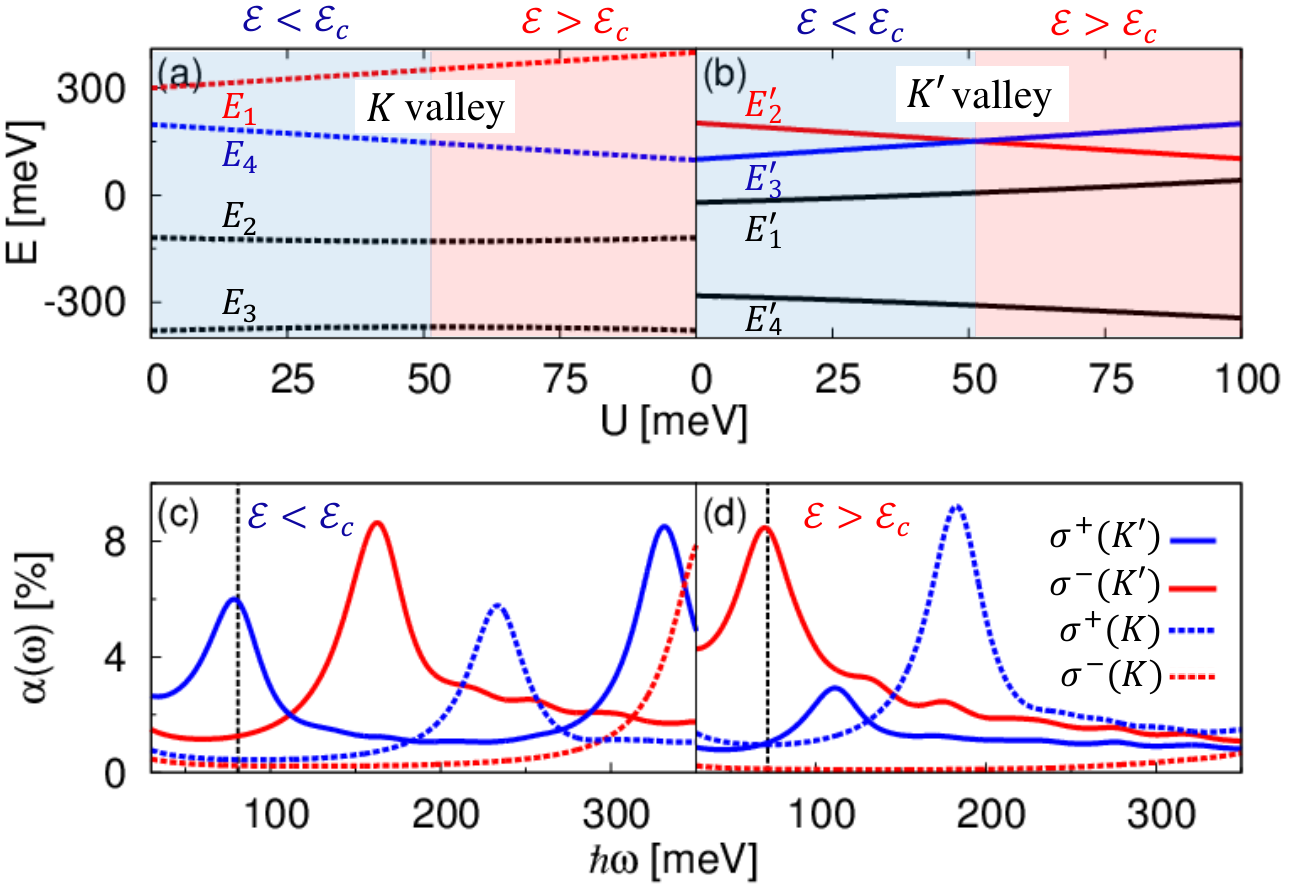}
\caption{Band-edge energies as a function of the staggered potential at the (a) $K$ and (b) $K'$ valleys. Helicity-resolved absorption spectra for both valleys when (c) $\mathcal{E}<\mathcal{E}_c$ ($U=10$ meV) and (d) $\mathcal{E}>\mathcal{E}_c$ ($U=80$ meV), dashed vertical: respective $1s$ exciton energies at 81 and 72 meV. At $K'$, the band crossing leads to a change of the winding number, and thus to a  change in helicity for the lowest $1s$ exciton absorption peak. }
\label{fig:EnergiesAbsorption}
\vspace{-0.3cm}
\end{figure}

We consider the setup in  Fig.~\ref{fig:Scheme}(a), a group-V ML deposited on a substrate magnetized along $\hat{\bm{z}}$, with an applied out-of-plane electric field, $\mathcal{E}$.  The ML contains A, B sublattices and $K$/$K'$ valleys, at which a low-energy effective Hamiltonian can be constructed. The Hilbert space  $\{ | \phi_{A,\tau}^{\uparrow}\rangle,  |\phi_{B,\tau}^{\uparrow}\rangle,  |\phi_{A,\tau}^{\downarrow}\rangle, |\phi_{B,\tau}^{\downarrow}  \rangle\}$ is given with basis functions using $p_{x,y}$ orbitals $|\phi_{A, \tau} \rangle=|- i \tau p_x^A  + p_y^A \rangle $, 
$|\phi_{B, \tau} \rangle =| i \tau p_x^B + p_y^B \rangle$, and $\tau=\pm 1$ are $K$/$K'$ valley indices. The effective Hamiltonian, $H=H_0 + H_{ex} + H_U + H_R$, has contributions  from the group-V ML, exchange, staggered potential, and Rashba spin-orbit coupling (SOC)~\cite{Dominguez2018:PRB}. 
In $H_0$ for the planar ML~\cite{Zhou2015:NL,Zhou2016:PRB,Zhou2018:npjQM,Song2014:NPGAM},
\begin{eqnarray}
H_0= \hbar v_F(  k_x \sigma_x \tau_z+k_y \sigma_y )  + \lambda_{SO} \sigma_z S_z \tau_z,
\end{eqnarray}
$v_F$ is the Fermi velocity, $k_{x,y}$ are momenta measured from $K$/$K'$, $\lambda_{SO}$ is the effective on-site SOC, $\sigma_i$,  $\tau_z$ and $S_z$ denote Pauli matrices for the orbital, valley, and spin degrees of freedom, respectively. A staggered exchange field induced in the ML arises from the magnetic proximity effect or atomic doping and leads to a term,
\begin{equation}
H_{ex}= M_A S_z (1+\sigma_z)/2+ M_B S_z (1-\sigma_z)/2,
\label{eq:ex}
\end{equation}
\vspace{0.05cm}

\noindent{where} $M_A$ ($M_B$) is the exchange field for the A (B) sublattice. The magnitude of the staggered potential due to the broken inversion symmetry in the ML, $H_U = U \sigma_z$ can be controlled by the electric field, while Rashba SOC is $H_R=3\lambda_R(\sigma_x S_y \tau_z-\sigma_y S_x)$, with its strength 
$\lambda_R$~\cite{Dominguez2018:PRB}. 

After a unitary transformation, $H$ becomes
\begin{widetext}
\begin{eqnarray}
\tilde H_{\tau}  = 
\begin{bmatrix}
E_1^{\tau}   &  \tau \cos \theta_{\tau} \hbar v_F k_{-\tau}   &  -i \sin \theta_{\tau} \hbar v_F   k_- & 0 \\
 \tau \cos \theta_{\tau} \hbar v_F k_{+\tau}     &  E_2^{\tau}     & 0 & -i\sin \theta_{\tau} \hbar v_F k_- \\
i \sin \theta_{\tau} \hbar v_F   k_+  & 0 &  E_3^{\tau}   &-\tau \cos \theta_{\tau} \hbar  v_F  k_{-\tau} \\
0  & i\sin \theta_{\tau} \hbar v_F k_+ &-\tau \cos \theta_{\tau} \hbar  v_F  k_{+\tau} & E_4^{\tau}
\end{bmatrix},  
\end{eqnarray}
\end{widetext}
where $\theta_{\tau}=(1/2)\cos^{-1} \left[(\bar M - \tau U)/\sqrt{(\bar M - \tau U)^2+36\lambda_R^2}\right]$, $k_{\pm} = k_x \pm i k_y$, $k_{\pm\tau} = k_x \pm i \tau k_y$. The conduction and valence band (CB, VB) edges at $K$ are $E_{1, 4} = \lambda_{SO} \pm U \pm M_{A, B}$, $E_{2, 3} = - \lambda_{SO}- \Delta M \pm \sqrt{(U-\bar M)^2+36\lambda_R^2}$, while at $K'$ they are $E_{2, 3}'=\lambda_{SO}\mp U \pm M_{B, A}$, $E_{1,4}'$ $= - \lambda_{SO}+\Delta M \pm\sqrt{(U+\bar M )^2+36\lambda_R^2}$, with $\bar M = (M_A+M_B)/2$ and $\Delta M = (M_A-M_B)/2$~\cite{Note:SM}. 

Figures~2(a) and (b) reveal that this explicit dependence of the band-edge energies on the staggered potential is inequivalent in the two valleys, such that we can reach a critical value $U_c=\bar M$ for the CB crossing at $K'$, just as depicted in Fig.~1(b). This valley asymmetry is the result of the nonzero $U$ and  $\Delta M$, where $U$ can be directly controlled by the $\mathcal{E}$-field: at $\mathcal{E}=\mathcal{E}_c$ there is CB reversal at $K'$ for electrical control of the winding number, as further discussed in Ref.~\cite{Note:SM}.

Since the pair of bands with smallest gap has the largest contribution to the lowest-energy exciton, we apply L\"{o}wdin's perturbation theory~\cite{Winkler:2003} and downfold $\tilde{H}_{\tau}$ to the lower CB and upper VB. In the regimes (I): $0<U<\bar M$, and (II):   $U>\bar M$, the downfolded $\tilde{H}_{\tau}$ at the $K'$ valley, up to the linear order in $k$, is given by
\begin{equation}
\tilde H_{K'}^{(I)} = \begin{bmatrix}
E_3'     &  i \sin \theta_+ \hbar v_F k_+  \\
-i \sin \theta_+ \hbar v_F k_-  &   E_1'
\end{bmatrix},  \\
\end{equation}
\begin{equation}
\tilde H_{K'}^{(II)} =
\begin{bmatrix}
E_2'     &  -\cos \theta'_+ \hbar v_F k_-  \\
-\cos \theta_+' \hbar v_F k_+  &   E_1'
\end{bmatrix},
\end{equation}
revealing the CB reversal at $U_c =\bar M$. In contrast, there is no CB reversal at the $K$ valley, 
\begin{equation}
\tilde H_K^{(I)}=
\begin{bmatrix}
E_4  &  i \sin \theta_{-} \hbar v_F  k_+   \\
- i \sin \theta_{-} \hbar v_F  k_-  & E_2
\end{bmatrix},
\end{equation}
such that $\tilde{H}_K^{(II)}$ has the same form as $\tilde{H}_K^{(I)}$.

The downfolded $\tilde{H}_{\tau}$ has the same form~\cite{Note:SM} as the chiral fermion model~\cite{Zhang2018:PRL} and we can readily identify the corresponding winding numbers, $w$. Unlike the unchanged $K$-valley winding number, the change of the band topology at the $K'$ valley in regimes (I) and (II) results both in a change of the winding number and the excitonic optical selection rules. Specifically, the excitons with angular quantum number $m$ are bright (optically allowed) for helicity $\sigma^{\pm}$ if $m=w\mp 1 + n N$~\cite{Zhang2018:PRL, Cao2018:PRL}, where $n$ is an integer and $N$ the degree of discrete rotational symmetry of the crystal lattice. Therefore, the change of winding number from $w=1$ to $-1$ in the $K'$ valley transforms the series of bright excitons with respect to the helicity of light. For example, the $1s$ exciton ($m=0$) is bright under $\sigma^+$ when $w=1$; if $w=-1$, it is instead bright under $\sigma^-$. 

So far, we have performed a single-particle analysis, from which implications on the exciton optical selection are made. Below, we verify these predictions
and take into account many-body effects by numerically solving the Bethe-Salpeter equation (BSE)~\cite{Rohlfing2000:PRB,Wu2015:PRB,Scharf2016:PRB,Note:SM,Scharf2017:PRL} to explain experimentally detectable findings,
\begin{equation}
[\Omega^{\tau}_{S} -\epsilon^\tau_c(\vec{k})+\epsilon^\tau_v(\vec{k})] \mathcal{A}^{S\tau}_{vc\vec{k}} =\sum\limits_{v'c'\vec{k}'}\mathcal{K}^{\tau}_{vc\vec{k},v'c'\vec{k}'}\mathcal{A}^{S\tau}_{v'c'\vec{k}'},
\label{eq:BSE}
\end{equation}
where $\Omega^{\tau}_{S}$ is the energy of the exciton state $\ket{\Psi^{\tau}_{S}}=\sum_{vc\vec{k}}\mathcal{A}^{S\tau}_{vc\vec{k}}\hat{c}^\dagger_{\tau c\vec{k}}\hat{c}_{\tau v\vec{k}}\ket{\mathrm{GS}}$~\cite{footnote:BSE} with the coefficients $\mathcal{A}^{S\tau}_{vc\vec{k}}$, the creation (annihilation) operator of an electron in a CB $c$ (VB $v$) $\hat{c}^\dagger_{\tau c\vec{k}}$ ($\hat{c}_{\tau v\vec{k}}$) in the valley $\tau$, and the ground state $\ket{\mathrm{GS}}$ with fully occupied VBs and unoccupied CBs. $\epsilon^\tau_n(\vec{k})$ with the band index $n=c(v)$ describes the eigenenergies of the single-particle Hamiltonian, $H_{\tau}  \eta^{\tau}_{n\vec{k}}=\epsilon^\tau_n(\vec{k}) \eta^{\tau}_{n\vec{k}}$, with the eigenstates $\eta^{\tau}_{n\vec{k}}$. The interaction kernel $\mathcal{K}^{\tau}_{vc\vec{k},v'c'\vec{k}'}$ accounts for the Coulomb interaction between electrons in the ML, determined from the dielectric environment~\cite{Keldysh1979:JETP,Cudazzo2011:PRB,Scharf2019:JPCM,Note:SM}. This framework also allows for an inclusion of the magnetic proximity effects on excitons~\cite{Scharf2017:PRL}, which can be crucial, but are absent from the common single-particle description.

The optical response is described by the absorption  
\begin{equation}
\alpha(\omega)=\frac{4e^2\pi^2}{c \: \omega}\frac{1}{A}\sum\limits_{S\tau}\left|\sum\limits_{vc\vec{k}} v_\pm (\vec k) \mathcal{A}^{S\tau}_{vc\vec{k}}\right|^2\delta(\hbar\omega-\Omega^{\tau}_{S}),
\label{eq:absorption}
\end{equation}
where $\omega$ and $c$ are the frequency and speed of light, $A$ the 2D unit area, and the velocity matrix elements are $v_{\pm} (\vec k) = \langle v \vec k | \hat v_{\pm} | c \vec k \rangle$ for helicities $\sigma^{\pm}$, with $ \hat v_{\pm} = \hat v_x \pm i \hat v_y$, and $\hat v_{x, y} = \partial H/\partial(\hbar  k_{x, y})$. The absorption in Figs.~\ref{fig:EnergiesAbsorption}(c) and (d) confirms a striking gate-controlled helicity reversal, anticipated from the single-particle picture. Remarkably, by  changing $\mathcal{E}<\mathcal{E}_c$ (at $U=10$ meV) to $\mathcal{E}>\mathcal{E}_c$ (at $U=80$ meV), 
we see a helicity reversal of the lowest $1s$ exciton peak from $\sigma^+$ to $\sigma^-$, using realistic parameters discussed further in Ref.~\cite{Note:SM}. Alternatively, the photoluminescence ($\sigma^+$ or $\sigma^-$), dominated by the $1s$ exciton due to relaxation, is also gate-controllable [Fig.~\ref{fig:Scheme}(a)], reflecting the changing band topology. 
 
\begin{figure}[t]
\centering
\includegraphics*[width=8.6cm]{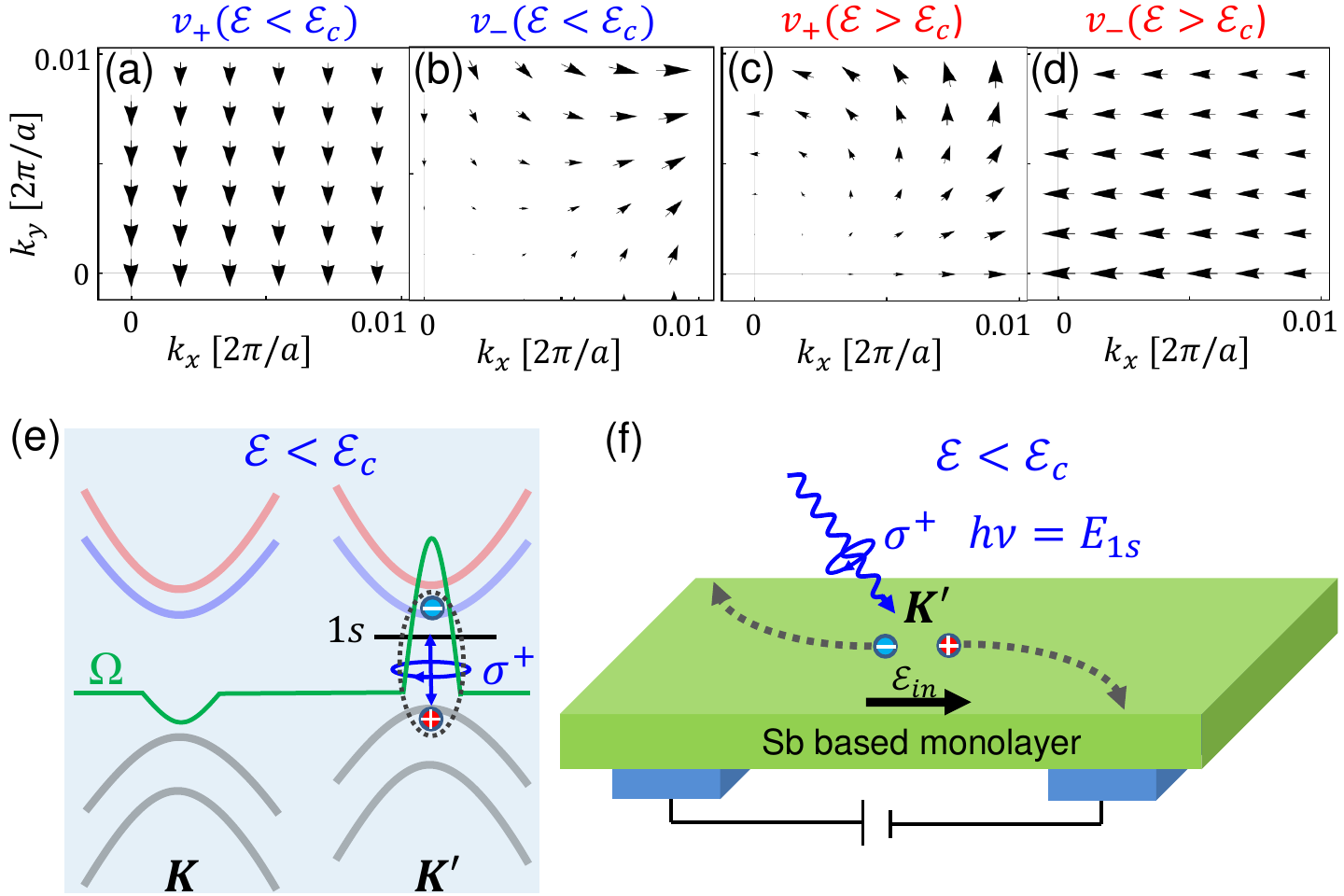}
\vspace{-0.5cm}
\caption{Winding patterns of the interband matrix elements of the velocity operators $v_{\pm} (\vec k)$ at the $K'$ valley with the $K'$ point at the origin.  
The real (imaginary) part of $v_{\pm}({\vec k})$ is given along the horizontal (vertical) direction. (a) $v_+ (\vec k)$,  
(b) $v_- (\vec k)$ for $\mathcal{E}<\mathcal{E}_c$ and   (c) $v_+ (\vec k)$, 
(d) $v_- (\vec k)$ for $\mathcal{E}>\mathcal{E}_c$. (e) Resonant optical excitation of $1s$ excitons with $\sigma^+$ light when 
$\mathcal{E}<\mathcal{E}_c$ and Berry curvature $\Omega \neq 0$ at the valleys. (f) Anomalous valley Hall effect with an applied  in-plane  
electric field, $\mathcal{E}_{in}$.}
\label{fig:Texture}
\vspace{-0.4cm}
\end{figure}

To illustrate the $\vec k$-space band topology, in Figs.~\ref{fig:Texture}(a)-(d) we plot the interband matrix elements of  $\hat v_{\pm} (\vec k)$ 
at the $K'$ valley~\cite{Note:SM}. Each complex matrix element is visualized as a vector with its origin located at  $\vec k$. These matrix elements are not physical observables and have an arbitrary phase, just as the eigenstates $|v \vec k \rangle$ and $|c \vec k \rangle$. To avoid such phase ambiguity and obtain meaningful patterns, we chose a specific gauge~\cite{Note:SM}. 

The change of winding number from $w=1$ ($\mathcal{E}<\mathcal{E}_c$) to $w=-1$  ($\mathcal{E}>\mathcal{E}_c$) is accompanied by a change of winding patterns, which are also consistent with the analytical analysis of the two-band model: $v_{\pm} (\vec k) \propto \exp(-i (w\mp 1) \phi_{\vec k})$, where $\phi_{\vec k}$ is the angle between $\vec{k}$ and the $x$-axis \cite{Zhang2018:PRL}. Since the system has a rotational symmetry ignoring crystal field effects, the envelope function of an exciton can be formally written as $\mathcal{A}_{vc\vec{k}} = f_m (|\vec k|) e^{i m \phi_{\vec k}}$, containing a radial part $f_m (|\vec k|)$ and an angular part with an angular quantum number $m$. Thus, the oscillator strength of the optical transition to an exciton state with angular quantum number $m$ can be expressed as,
\begin{eqnarray}
O_{\pm}^m \propto \left| \sum_{\vec k} f_m (|\vec k|) e^{i m \phi_{\vec k}} v_{\pm} (\vec k) \right|^2, 
\end{eqnarray}
which determines the contribution of this particular excitation to the absorption in Eq.~(\ref{eq:absorption}). To have a nonzero integral and thus a bright exciton, the angular part of the integral has to vanish, i.e., the phase winding of the velocity matrix element has to be canceled by the angular part of the exciton envelope function. Since $f_m (|\vec k|)$ is localized around $\vec k =0$ (Wannier excitons), we can estimate the oscillator strength and thus the brightness of a specific exciton state with angular quantum number $m$ based on the winding patterns of $v_{\pm} (\vec k)$. 

When $\mathcal{E}<\mathcal{E}_c$, $v_+(\vec k)$ is nearly uniform in the vicinity of the $K'$ point ($\vec k =0$). Therefore, the angular part of the integral vanishes for $m=0$, which means that $s$ excitons are bright under $\sigma^+$ light. However, $v_-(\vec k)$ winds clockwise twice around the $K'$ point through any counter clockwise contour enclosing the $K'$ point, which implies bright $d$ excitons ($m=2$) under $\sigma^-$ light. In contrast, when $\mathcal{E}>\mathcal{E}_c$, $v_-(\vec k)$ has uniform phases and therefore supports bright $s$ excitons ($m=0$) under $\sigma^-$. Additionally, $v_+(\vec k)$ winds counterclockwise twice around the $K'$ point, indicating bright $d$ excitons ($m=-2$) under $\sigma^+$. These findings agree with our single-particle analysis as well as the BSE results in Figs.~\ref{fig:EnergiesAbsorption}(c) and (d).
 
A tunable band topology and the resulting helicity reversal can also be probed through transport. In Fig.~\ref{fig:Texture}(e) we
show how, at $K'$,  the helicity-selective optical excitations 
($\sigma^+$ for $\mathcal{E}<\mathcal{E}_c$ and $\sigma^-$ for $\mathcal{E}>\mathcal{E}_c$~\cite{Note:SM}) near $1s$ exciton resonance yield a 
carrier density imbalance in the two valleys. Consequently, under an in-plane electric field, such photoexcited carriers can be driven to the transverse direction by the nonzero Berry curvature, $\Omega$,
giving rise to an anomalous valley Hall effect in Fig.~\ref{fig:Texture}(f).

To explore the experimental feasibility of our proposal we summarize its requirements~\cite{Note:SM}:
(1) Direct-gap chiral fermions in the valleys, (2) Exchange field to break  time-reversal symmetry to induce valley splitting, 
(3) Staggered potential to break inversion symmetry, which is also controllable by an electric field.
We use first-principles calculations and seek realistic materials. Recently, ML Sb has been successfully synthesized and characterized~\cite{Ji2016:NC,Shao2018:NL}. Fully hydrogenated ML Sb (SbH), important for our proposal~\cite{Note:SM}, was found to be a 2D quantum spin Hall insulator with a giant direct topological gap of 410 meV, while retaining a stable structure up to 400 K, revealing its potential as a platform for room-temperature quantum devices~\cite{Song2014:NPGAM}. The desired exchange field and staggered potential in our model can be induced to ML SbH either through proximity effects from magnetic substrates (e.g., LaFeO$_3$)~\cite{Zhou2016:PRB}  or doping of magnetic atoms (W, Cr, and Mo)~\cite{Zhou2015:NL}. 

\begin{figure}[t]
\centering
\includegraphics*[width=8.6cm]{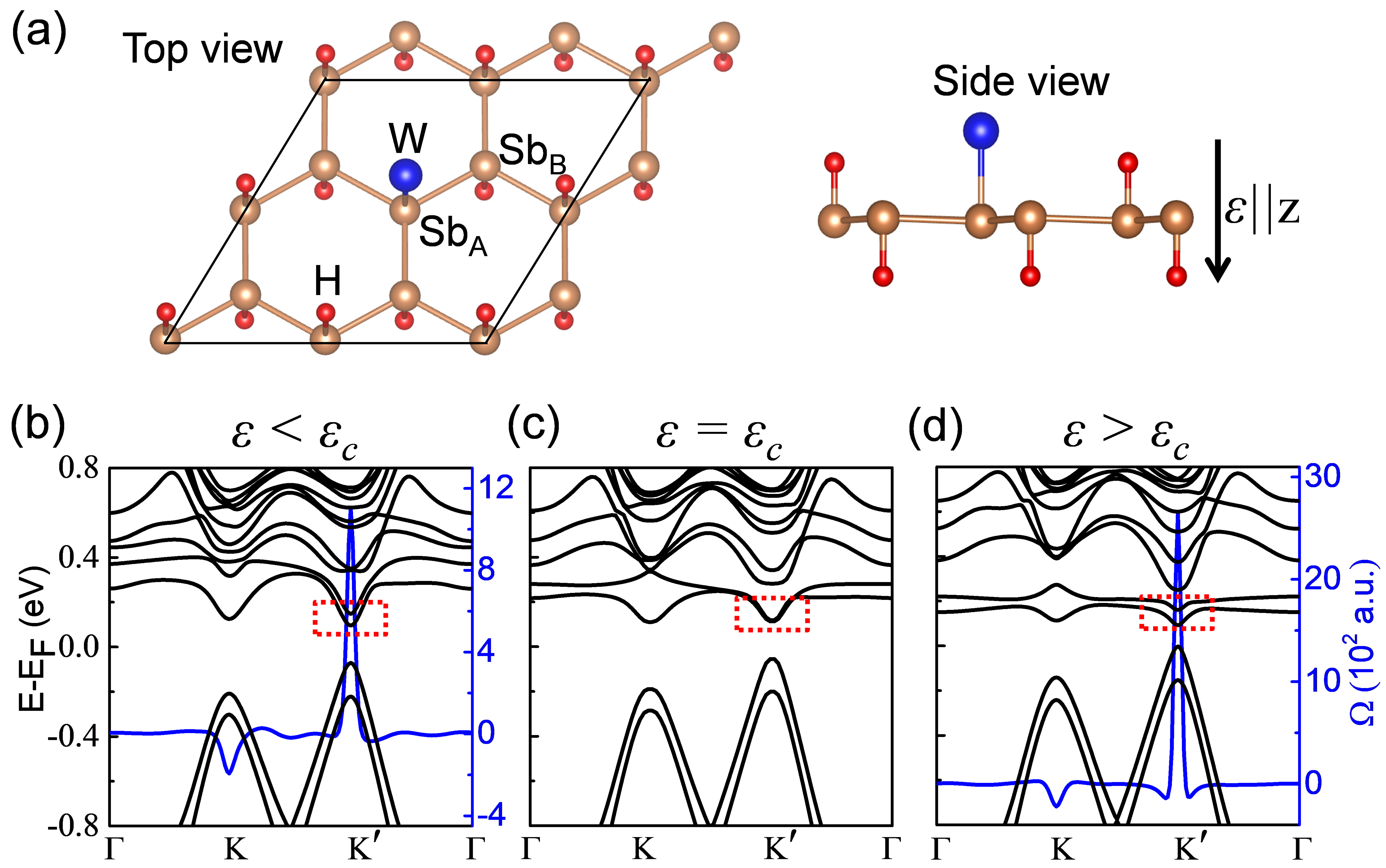}
\vspace{-0.5cm}
\caption{(a) Schematic of the ML SbH doped with one W atom 
and the 2$\times$2 supercell (black lines) used in our calculations.
A related electronic structure and Berry curvature, $\Omega$, for the electrical field, $\mathcal{E}$: 
(b) smaller ($0$ V/nm), (c) equal ($2$ V/nm), and (d) greater ($4$ V/nm) than the critical value $\mathcal{E}_c$, 
displaying a CB inversion  in the $K'$ valley (highlighted in
rectangles).}
\label{fig:Structure}
\vspace{-0.3cm}
\end{figure}

A fully relaxed W-doped ML SbH hexagonal structure is shown in Fig.~\ref{fig:Structure}(a) and its experimental feasibility is discussed in Ref.~\cite{Note:SM}. To demonstrate the manipulation of band topology by the electric field, we show in Figs.~\ref{fig:Structure}(b)-(d) first-principles calculations for the evolution of the electronic structure and $\Omega$, supporting the scenario from Fig.~\ref{fig:Texture}(e)~\cite{Note:SM}. At $\mathcal{E}=0$,  Fig.~\ref{fig:Structure}(b) reveals a desirable direct topological gap $\sim 180$ meV. For $\mathcal{E}$ applied along  the $\hat{\bm{z}}$-direction, the gap between the two CBs at $K'$ reduces and closes at $\mathcal{E}= 2$ V/nm [Fig.~\ref{fig:Structure}(c)]. For even larger $\mathcal{E}$~\cite{note:gate} in Fig.~\ref{fig:Structure}(d), such a gap between the two CBs reopens, leading to a change of the $K'$ winding number. The corresponding band inversion and change of band topology indicate that W-doped ML SbH is indeed a good candidate to observe our predictions. This is further corroborated by the
resolution of helicity-sensitive detection~\cite{Cerne2003:RSI,Ellis2013:SR,Wang2017:NL,Kumar2016:O} and studied robustness to disorder~\cite{Note:SM}, 
while changes of the W concentration or strain offer additional tunability~\cite{Note:SM}.

To generalize considerations from this work, it would be interesting to examine how the influence of tunable band topology on optical selection rules can be extended to other materials systems, beyond chiral fermions.
 
While we have focused on magnetically-doped MLs, our predictions could instead employ magnetic proximity-modified excitonic transitions. This approach could also be studied using the Bethe-Salpeter equation~\cite{Scharf2017:PRL} for considering atomically-thin magnetic substrates~\cite{Zollner2019:PRB,Shayan2019:NL}. The advantage of magnetic proximity effects over applying a magnetic field to remove the valley degeneracy is their relative magnitude~\cite{Zutic2019:MT,Qi2015:PRB,Zhao2017:NN,Zhong2017:SA}, exceeding a modest Zeeman splitting from typical $g$-factors in MLs~\cite{Srivastava2015:NP,MacNeill2015:PRL,Stier2016:NC,Plechinger2016:NL}. Electrically-tunable magnetic proximity effects in MLs~\cite{Lazi2016:PRB,Xu2018:NC,Cortes2019:PRL,Zhao2019:2DM,Zollner2019:NJP,Tao2019:PRB} could reduce the critical electric field to alter the band topology. In particular, it would be desirable to implement a fast electrical reversal of helicity as an element of the suggested ML-based spin-lasers~\cite{Lee2014:APL}, which could exceed the performance of the best conventional counterparts~\cite{Lindemann2019:N}, without spin-polarized carriers. 

\begin{acknowledgments}
We thank J. Cerne, R. K. Kawakami, and P. Vora for discussion about experimental implementation of our work. This work was mainly supported by the U.S. DOE, Office of Science BES, Award No. DE-SC0004890. B.S. was supported by the German Research Foundation  (DFG) through SFB 1170 Project-ID 258499086 and the W\"urzburg-Dresden Cluster of Excellence on Complexity and Topology in Quantum Matter -- \textit{ct.qmat} (EXC 2147, Project-ID 39085490) and by the ENB Graduate School on Topological Insulators. Computational work was supported by the UB CCR.
\end{acknowledgments}

\end{document}